\documentclass[%
 aip,
 amsmath,amssymb,
 reprint,%
]{revtex4-1}

\usepackage{graphicx}
\usepackage{dcolumn}
\usepackage{bm}

\usepackage[utf8]{inputenc}
\usepackage[T1]{fontenc}
\usepackage{etoolbox}

\usepackage{multirow}
\usepackage{physics}
\usepackage{xcolor}
\usepackage{graphicx}
\usepackage{dcolumn}
\usepackage{bm}

\makeatletter
\def\@email#1#2{%
 \endgroup
 \patchcmd{\titleblock@produce}
  {\frontmatter@RRAPformat}
  {\frontmatter@RRAPformat{\produce@RRAP{*#1\href{mailto:#2}{#2}}}\frontmatter@RRAPformat}
  {}{}
}%
\makeatother
\begin{document}


\title[]{Multi-objective optimization for retinal photoisomerization models with respect to experimental observables}

\author{Rodrigo A. Vargas-Hern\'andez}
\email{r.vargashernandez@utoronto.ca}
\affiliation{Chemical  Physics  Theory  Group,  Department  of  Chemistry, University of Toronto,Toronto, Ontario, M5S 3H6, Canada} 
\affiliation{Vector Institute, Toronto, Ontario.}

\author{Chern Chuang}
\affiliation{Chemical  Physics  Theory  Group,  Department  of  Chemistry, University of Toronto,Toronto, Ontario, M5S 3H6, Canada}

\author{Paul Brumer}
\affiliation{Chemical  Physics  Theory  Group,  Department  of  Chemistry, University of Toronto,Toronto, Ontario, M5S 3H6, Canada} 

\date{\today}

\begin{abstract}
The fitting of physical models is often done only using a single target observable. However, when multiple targets are considered, the fitting procedure becomes cumbersome, there being no easy way to quantify the robustness of the model for all different observables. Here, we illustrate that one can  jointly search for the best model for each desired observable through multi-objective optimization. To do so we construct the Pareto front to study if there exists a set of parameters of the model that can jointly describe multiple, or all, observables. To alleviate the computational cost, the predicted error for each targeted objective is approximated with a Gaussian process model, as it is commonly done in the Bayesian optimization framework. We applied this methodology to improve three different models used in the simulation of stationary state $cis-trans$  photoisomerization of retinal in rhodopsin. Optimization was done with respect to different experimental measurements, including emission spectra, peak absorption frequencies for the $cis$ and $trans$ conformers, and the energy storage. 
\end{abstract}

\maketitle

\begin{quotation}

\end{quotation}

\section{\label{sec:Intro} Introduction}

In chemical physics, observables such as absorption spectra are often simulated with closed-form model Hamiltonians or Liouvillians. Examples studied in detail here are models of $cis-trans$ photoisomerization of retinal in rhodopsin.
The parameters of such models, $\boldsymbol{\theta}$, are usually fit by minimizing a scalar error function, $f(\boldsymbol{\theta}):$  
$\mathbb{R}^{n} \to \mathbb{R}$, that quantifies the accuracy of the model with respect to a target set of observables.
Optimization of $f(\boldsymbol{\theta})$ can be achieved by gradient-based methods or by sampling algorithms that self-learn from previous iterations, e.g., genetic algorithms \cite{GeneticAlg} and Bayesian optimization (BO) \cite{BO_Adams,BO_Freitas}.
For observables that depend on quantum dynamics calculations, gradient-based methods are rarely used, due to inaccuracies in approximating the gradient of $f(\boldsymbol{\theta})$ with respect to the free parameters of the model Hamiltonian or Liouvillian.  
By contrast, sampling algorithms have proven to be robust search tools in chemical physics. 
For example, BO has been used to optimize density functionals,\cite{BO_DFT_Reiher,BO_DFT} generate low energy molecular conformers,  \cite{BO_PES,BO_geo} and inverse design potential energy surfaces for reactive molecular systems \cite{ravh_njp}. BO has also been successfully applied for screening chemical compounds \cite{BO_mat,BO_dft_calc,BO_phonono_transport}, minimizing the energy of the Ising model \cite{BO_ising}, and has recently been used to optimize laser pulses for molecular control \cite{RK_MLQDynamics,BO_RK} .

Many efforts have been devoted to mapping the search for a physical model into an optimization problem. 
However, in chemical physics comparing different models with respect to a single observable can lead to overfitting. To account for more realistic physical models the search-space must include multiple observables.
One common strategy is to adopt certain linear combinations of the error functions that account for the discrepancy, $f_i$, between the model's prediction and that of each of the target observables, 
\begin{eqnarray}
{\cal L}(\boldsymbol{\theta}) = \sum_i^{M}  w_i f_i(\boldsymbol{\theta}).
\label{eqn:linear_loss}
\end{eqnarray}
However, mapping the search of a physical model using Eq. \ref{eqn:linear_loss} assumes that (i) the values of the coefficients, $w_i$'s, are known, and (ii) there exists a single optimal minimum that jointly describes all observables. Both are extremely strong assumptions. Furthermore, the possibility of having multiple local minima strongly depends on how the values of $w_i$ are chosen. 

Such difficulties can be circumvented by not assuming a scalar function (a certain set of $w_i$'s) to quantify the accuracy of the model but instead by working with a vector function whose elements are the individual error functions: $\mathbf{F}(\boldsymbol{\theta}) = \left [f_1(\boldsymbol{\theta}),\cdots,f_M(\boldsymbol{\theta})\right ]$.
Here, we propose that deeper insights can be gained about the model-space by learning the boundaries of $\mathbf{F}(\boldsymbol{\theta})$ at different values of $\boldsymbol{\theta}$ and by reformulating the problem in terms of a multi-objective optimization scheme.  
Multi-objective optimization has been used to guide the experimental search for materials with desired properties \cite{MOBO_materials_0,MOBO_materials_1,MOBO_materials_2} and to optimize chemical reactions. \cite{MOBO_aspuru,MOBOpt_chem_rxn} This procedure was also used to fit the potential energy surface for diatomic molecules from their rotational and vibrational spectra. \cite{JPR_GA}

In this paper we apply multi-objective optimization to examine, and improve, three proposed models of retinal photoisomerization, the first step in vision. The results identify advantages and disadvantages of each model.
The paper is organized as follows: Section \ref{sec:MOBOpt} provides an introduction to the central tool used in this work, an algorithm to construct the Pareto front.
In Section \ref{sec:models}, we apply the method to three different models for retinal photoisomerization.
Sections \ref{sec:results} and \ref{sec:summary} provide results and the discussion, respectively.

As we have previously shown \cite{Tscherbul2014JPCA,dodin2016,PBrumer_JPCL} the fact that nature operates with incoherent (solar)  radiation implies that we should focus on properties in the stationary state, the approach adopted below. 

\section{Multi-objective Optimization}
In this section we present a brief introduction to multi-objective optimization and to the concept of the Pareto front (PF) \cite{MOBO_book}.

\subsection{\label{sec:Pareto Front}Pareto Front}
Multi-objective optimization considers the optimization of multiple objective functions and is usually formulated as seeking a minimum of the vector function;
\begin{eqnarray}
\boldsymbol{\theta}^* = \text{arg min}_{\boldsymbol{\theta}} \; \mathbf{F}(\boldsymbol{\theta})  = \text{arg min}_{\boldsymbol{\theta}}  \; \left [f_1(\boldsymbol{\theta}), \cdots, f_M(\boldsymbol{\theta})\right ].
\label{eqn:mobopt_loss}
\end{eqnarray}
Since it is unknown if a single $\boldsymbol{\theta}$ could jointly minimize all $f_i$'s, we must consider all different values where components of $\mathbf{F}$ are minimum. 
To quantify the improvement of $\mathbf{F}$, we introduce the concept of \emph{Pareto dominant points} $\boldsymbol{\theta}_\ell$, defined as,  
\begin{enumerate}
\item $f_i(\boldsymbol{\theta}_\ell) \leq f_i(\boldsymbol{\theta}_k)$, for all target functions.
\item $f_i(\boldsymbol{\theta}_\ell) < f_i(\boldsymbol{\theta}_k)$, for at least a single target function.
\end{enumerate}

The PF $Y_{P} = \{\mathbf{F}(\boldsymbol{\theta}_1), \cdots,\mathbf{F}(\boldsymbol{\theta}_n) \}$ are those arising from all the dominant points, $X_{P} = \{\boldsymbol{\theta}_1,\cdots,\boldsymbol{\theta}_n\}$, where $n$ is the number of Pareto dominant points.\\

\begin{figure}[h!]
\centering
\includegraphics[width=0.3\textwidth]{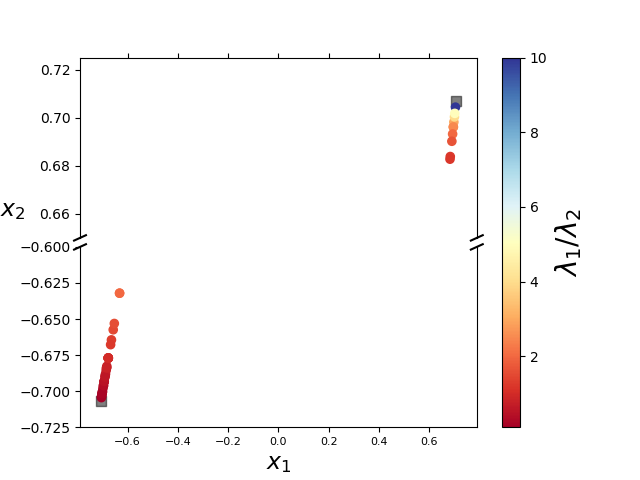}
\includegraphics[width=0.3\textwidth]{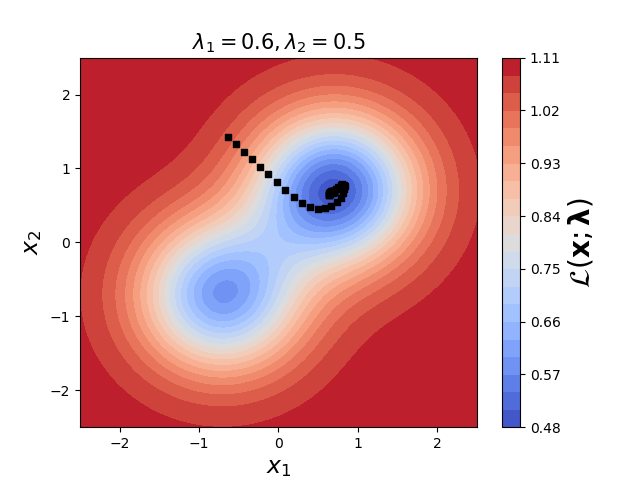}
\includegraphics[width=0.3\textwidth]{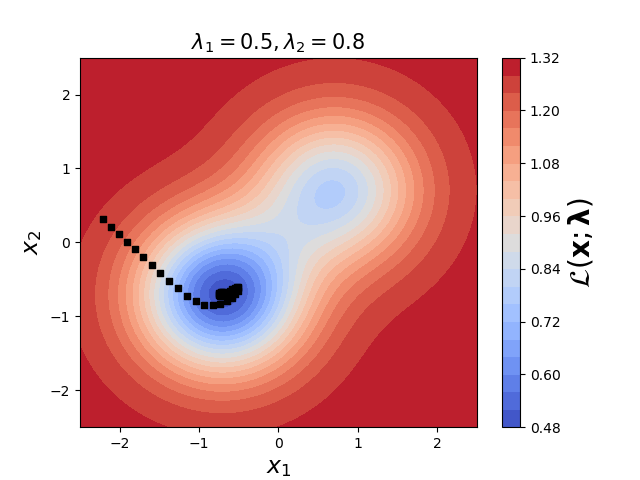}
\caption{
(upper panel) The symbols represent the minimizer of ${\cal L}_{FF}(\mathbf{x},\boldsymbol{\lambda}) = \sum_{i=1}^{2} \lambda_i f_i(\mathbf{x})$, as a function of  $\lambda_1/\lambda_2$, and the black-square symbols are the optimizers, $\mathbf{x}^* $, for each individual $f_i$ function.
(middle and lower panels) Function ${\cal L}_{FF}(\mathbf{x},\boldsymbol{\lambda})$,  for the sample (middle panel) $\lambda_1 = 0.6$ and $\lambda_2 = 0.5$, and (lower panel) $\lambda_1 = 0.5$ and $\lambda_2 = 0.8$.
The black symbols represent the optimization trajectory with gradient descent started with a random $\mathbf{x}$.
}
\label{fig:fonseca_l}
\end{figure}

\subsection{Example of Pareto Front}
As a pedagogical example of the PF consider the linear combination of the Fonseca-Fleming functions  \cite{fonseca_functions},
\begin{eqnarray}
{\cal L}_{FF}(\mathbf{x};\boldsymbol{\lambda}) = \lambda_{1} f_1(\mathbf{x}) + \lambda_{2} f_2(\mathbf{x}), \label{eqn:L_FF}
\end{eqnarray}
where $f_{1}(\mathbf{x}) =1- e^{- \sum_{i=1}^2\left(x_i -  \frac{1}{\sqrt{2}} \right )^2  }$ and $f_{2}(\mathbf{x}) =1- e^{- \sum_{i=1}^2\left(x_i +  \frac{1}{\sqrt{2}} \right )^2 }$. 
As is evident, the function ${\cal L}_{FF}$ also depends on the linear combination coefficients, $\boldsymbol{\lambda} = [ \lambda_{1}, \lambda_{2}]$, influencing the location of the minimizer; $\mathbf{x}^{*} = \arg\min_{\mathbf{x}}{\cal L}(\mathbf{x};\boldsymbol{\lambda})$.
$f_1$ and $f_2$  have different minimizers: for $f_1$ at $\mathbf{x}^*_1 = \left [\frac{1}{\sqrt{2}},\frac{1}{\sqrt{2}} \right]$, and for $f_2$ at $\mathbf{x}^*_2 = \left [-\frac{1}{\sqrt{2}},-\frac{1}{\sqrt{2}}\right]$.

\begin{figure}[h!]
\centering
\includegraphics[width=0.4\textwidth]{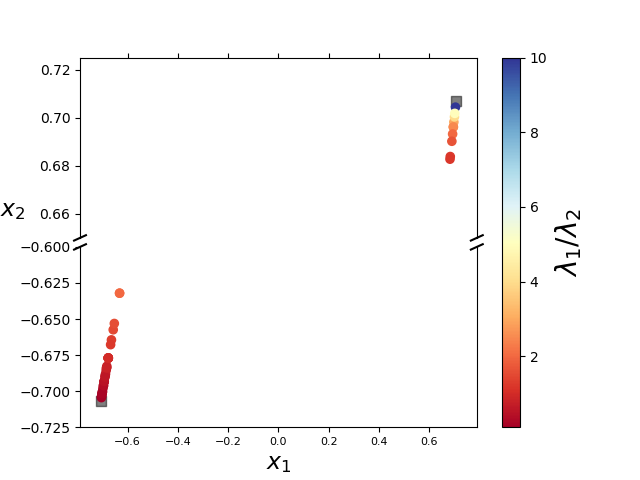}
\includegraphics[width=0.4\textwidth]{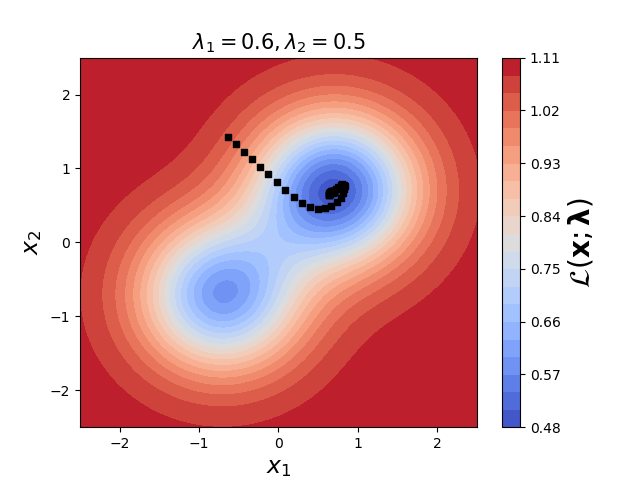}
\caption{The black markers represent the PF for the Fonseca-Fleming functions found with the MOBOpt algorithm. (upper panel) 
The blue markers are functional values of both $f_{1}$ and $f_{2}$ functions, and (lower panel) the $\mathbf{x}$ value for each Pareto point.}
\label{fig:fonseca_PF}
\end{figure}

From Figure \ref{fig:fonseca_l}, we can observe that by varying the values of $\lambda_{1}$ and $\lambda_{2}$, the location of $\mathbf{x}^*$ moves between $\mathbf{x}^*_1$ and $\mathbf{x}^*_2$, the minimizers of $f_1$ and $f_2$.
Tuning the values $\boldsymbol{\lambda}$ is not a trivial process, but we could study the optimization of multiple functions by means of multi-objective optimization, Eq. \ref{eqn:mobopt_loss}.
For the Fonseca-Fleming functions example, the Pareto optimal points connect both minima of the $f_1$ and $f_2$ functions, as these points describe where, in the function-space, $f_1$ and $f_2$ are minimum. 
Figure \ref{fig:fonseca_PF} depicts, in the upper panel, a set of points (blue) obtained by evaluating $[f_1(\mathbf{x}),f_2(\mathbf{x})]$ at various points $\mathbf{x}$. The Pareto dominant points $\mathbf{x}'$ generate a set of  $[f_1(\mathbf{x}'),f_2(\mathbf{x}')]$ that give the black markers in the upper panel, which form the PF. 
In principle, since we seek minimization, the PF is the relevant quantity in the upper panel.
However, the other (blue) points are shown to give a better picture of the $[f_1(\mathbf{x}),f_2(\mathbf{x})]$  for all $\mathbf{x}$.
The Pareto dominant points are shown as black markers in the lower panel.

The PF, for this example and those obtained for photoisomerization models, were found using the multi-objective Bayesian optimization (MOBOpt) algorithm, explained  below.
 
\subsection{\label{sec:MOBOpt} Optimization Algorithm}
One of the key advantages of machine-learning algorithms is the ability to approximate any function based on few data points. 
Here, we approximate each $f_i$ with an individual Gaussian process (GP), as is commonly done in single-objective BO,
\begin{eqnarray}
f_{i} \sim {\cal {GP}}_{i} (\mu, \Sigma),
\end{eqnarray}
where $\mu$ is the mean and $\Sigma$ is the covariance of the model. 
Gaussian process prediction for a new point $\mathbf{x^*}$ is achieved by,
\begin{eqnarray}
\mu(\mathbf{x^*}) &=& \mathbf{k}(\mathbf{x^*},\mathbf{X})^\top \left [ K(\mathbf{X},\mathbf{X}) + \sigma_n 1 \right]^{-1} \mathbf{y} \label{eqn:gp_mu}\\
\sigma(\mathbf{x^*}) &=& k(\mathbf{x^*},\mathbf{x^*}) \left [ K(\mathbf{X},\mathbf{X}) + \sigma_n 1 \right]^{-1} \mathbf{k}(\mathbf{x^*},\mathbf{X}),\label{eqn:gp_cov}
\end{eqnarray}
where $\mu(\mathbf{x^*})$ and $\sigma(\mathbf{x^*})$ are the mean and standard deviation of the posterior distribution. $\mathbf{X}$ and $\mathbf{y}$ are the training points, and $K(\cdot,\cdot)$ is the covariance matrix, whose matrix elements are computed using the kernel function, $K_{ij} = k(\mathbf{x}_i,\mathbf{x}_j)$.
The entries of $\mathbf{k}(\mathbf{x^*},\mathbf{X})$ are between $\mathbf{x^*}$ and all the training data $\mathbf{X}$.
For all calculations in this work we used the Matern 2/5 kernel,
\begin{eqnarray}
k_{Mat} (\boldsymbol{\theta}_i,\boldsymbol{\theta}_j) = \left [ 1 + \sqrt{5}r(\boldsymbol{\theta}_i,\boldsymbol{\theta}_j) + \frac{5}{3}r^2(\boldsymbol{\theta}_i,\boldsymbol{\theta}_j) \right]\exp^{-\sqrt{5}r(\boldsymbol{\theta}_i,\boldsymbol{\theta}_j)},
\end{eqnarray}
where $r^2(\boldsymbol{\theta}_i,\boldsymbol{\theta}_j) = (\boldsymbol{\theta}_i - \boldsymbol{\theta}_j)^\top M (\boldsymbol{\theta}_i - \boldsymbol{\theta}_j)$, and $M$ is a diagonal matrix that contains different length scale constants, $\{\ell_i\}_{i=1}^{d}$, for each dimension of $\boldsymbol{\theta}$.
Its dimension, $d$, is the number of parameters.
The optimal kernel parameters ($\boldsymbol{\theta}^*$) were found by maximizing the log-marginal likelihood, 
\begin{eqnarray}
\boldsymbol{\theta}^* &=& \arg\max_{\boldsymbol{\theta}} \;\; \log p(\mathbf{y}| \mathbf{X})  \nonumber  \\
&=& \arg\max_{\boldsymbol{\theta}} \left [ -\frac{1}{2} \mathbf{y}^\top K(\mathbf{X},\mathbf{X})^{-1}\mathbf{y} -\frac{1}{2}\log \left | K(\mathbf{X},\mathbf{X})  \right | - \frac{n}{2}\log{2\pi} \right ]. \nonumber  \\
\label{eqn:LML}
\end{eqnarray}
For more details about GPs we refer the reader to Ref. \cite{gpbook}. 

The second key component in BO is the acquisition function that guides the samples through the optimization towards the minimum of the scalar function under study. 
For multi-objective BO optimization, the acquisition function must describe how the PF change by sampling all the objective functions at different $\boldsymbol{\theta}$.
At each iteration, we used the manifold of $\{ {\cal GP}_{i} \}_{i = 1}^{M}$ to infer  $X_{P}(t)$ and $Y_{P}(t)$, where $t$ denotes the iteration number, using the non-dominated sorting genetic algorithm II (NSGAII) \cite{NSGA-II}.
The proposed samples are suggested by,
\begin{eqnarray}
\mathbf{x}_{next} = \arg\max \;\; \left [ q \left ( \frac{d_{\ell,f} - \mu_f}{\sigma_f} \right ) + (1-q)\left (\frac{d_{\ell,x} - \mu_x}{\sigma_x} \right) \right ],
\label{eqn:mobopt_aqn}
\end{eqnarray}
where $d_{\ell,f}$ is the L-2 norm between the point $\ell$ and all other $t$ points, both provided by the NSGA-II algorithm. $\mu_x$ and $\sigma_x$ are the average and standard deviation for all $d_{\ell,x}$ and $q$ is a hyperparameter associated with exploration-exploitation; $q\in [0,1]$ . If $q \approx 1$, we will sample more points close to the approximate PF (exploitation), and when $q \approx 0$ the points are sampled from the parameter space where the functions are minimal (exploration). For future references we denote this algorithm as the multi-objective Bayesian optimization (MOBOpt) algorithm. For more details consult Refs. \cite{MOBOpt,MOBOpt_git}.

\section{\label{sec:models} $cis-trans$ Isomerization Models}
Here we introduce the physical models used to simulate the emission spectra and the maximum absorption frequencies for the $cis$ and $trans$ retinal conformers, and the energy storage, the energy difference between these two conformers in rhodopsin.
We then focus on the extent to which these models can represent observed data.

\subsection{\label{sec:STmodels} Single and two-mode models}
The numerical simulation of minimal models has been a fruitful way of studying  \emph{11-cis}$\to$all-\emph{trans}   photoisomerization of retinal in rhodopsin in the last decades. 
Some of the most well known models rely on a parametrized form of the two lowest  electronic energy states to describe the isomerization \cite{TSTM_1994,TSTM_1995,Stock_2000_JCP,Stock_2000_CP}. The simplest model uses a rotational coordinate $\phi$ and  potentials
\begin{eqnarray}
V^{SM}_{0,0} &=& \frac{1}{2} W_0(1-\cos\phi)\\
V^{SM}_{1,1} &=& E_1 - \frac{1}{2} W_1(1-\cos\phi),
\label{eqn:SM_pes}
\end{eqnarray}
where $V^{SM}_{i,i}$ are the diabatic potential energy surfaces (PESs) for the ground ($i=0$) and excited ($i=1$) electronic states. 
The coupling between different electronic states, $V_{i,j}$, is assumed to be constant, $V_{0,1} = V_{1,0} = \lambda$. 
For this model, the free parameters to be optimized are $\boldsymbol{\theta}_{SM} = [E_1,W_0,W_1,\lambda]$.
We refer to this model as the \emph{single-mode} ($SM$) model.
By including an additional vibrational degree of freedom, $x$, each diabatic PES now includes the terms\cite{TSTM_1994,TSTM_1995,Stock_2000_JCP,Stock_2000_CP},
\begin{eqnarray}
V^{TM}_{0,0} &=&  V^{SM}_{0,0} + \frac{\omega}{2}x^2 \\
V^{TM}_{1,1} &=&  V^{SM}_{1,1} + \frac{\omega}{2}x^2  + \kappa x.
\label{eqn:TM_pes}
\end{eqnarray}
We refer to this model as the \emph{two-mode} ($TM$) model, and the coupling between the ground and excited state is defined as $V^{TM}_{0,1} = V^{TM} _{1,0} = \lambda x$.
The free parameters in this model are $\boldsymbol{\theta}_{TM} = [E_1,W_0,W_1,\kappa,\lambda]$. For all the calculations presented here, we fixed the value of $\omega$ to 0.19 eV.

We also consider a third model inspired by (but not the same as) the two-state-three-mode model reported in Ref. \cite{Olivucci_model}, where
\begin{eqnarray} 
V^{Omod}_{0,0} &=& \frac{\omega}{2}x^2 - W_0 \cos\phi - W_1\cos(2\phi)\\
V^{Omod}_{1,1} &=& E_1 + \frac{\omega}{2}x^2 + \kappa x - W_2\cos\phi  + W_3(1-\cos(2\phi))\\
V^{Omod}_{0,1} &=& V^{Omod}_{1,0} = \lambda \sin(2\phi).
\label{eqn:Olivucci_pes}
\end{eqnarray}
Here $V^{Omod}_{i,i}$ are the diabatic PESs for the ground and excited states, and $V^{Omod}_{0,1}$ is the coupling between the two states.
The free parameters of this model are  $\boldsymbol{\theta}_{Omod} = [\omega,W_0,W_1,E_1,\kappa,W_2,W_3,\lambda]$.

The kinetic energy operators of these models are written as $T=-\frac{1}{2m}\frac{\partial^2}{\partial\phi^2}+\frac{\omega}{2}\frac{\partial^2}{\partial x^2}$, except for the $SM$ model where only the first term exists. Here, we take $\hbar=1$ and the total Hamiltonian is $H_\mathrm{s}=V+T$.
The parameters of all three models as reported in the literature are given in Table \ref{table:models}.

In the following sections we present the PF for these three models with respect to different experimental measurements.
They are three emission spectra excited at different wavelengths, the maximum absorption frequencies for the $cis$ and $trans$ conformers, and energy storage. The energy storage is defined as the energy difference between the lowest state in the \textit{trans-}well ($\frac{\pi}{2}\le\phi<\frac{3\pi}{2}$) and that in the \textit{cis-}well ($-\frac{\pi}{2}\le\phi<-\frac{\pi}{2}$). These two states are also instrumental in defining the absorption spectra of the \textit{cis}- and \textit{trans}-conformers, while the specific definitions of the other five spectroscopic observables are given in the next section.

The aim of this analysis is twofold; first to compare the best set of parameters for each model, and second to determine the extent to which the models can agree with experimental data. We emphasize, once again, that our interest lies in the stationary state that models natural excitation with sunlight \cite{Tscherbul2014JPCA,dodin2016,PBrumer_JPCL}.  All results were computed with the MOBOpt algorithm described in Section \ref{sec:MOBOpt}.\\

\begin{table}[h]
 \caption{The literature value of the system parameters: $SM$ and $TM$ models are from Refs. \cite{TSTM_1994,TSTM_1995,Stock_2000_JCP,Stock_2000_CP}, and  $Omod$ model is based on Ref. \cite{Olivucci_model}. All parameters are  given in [eV].}
\begin{tabular}{c c c | c c }
                 & $SM$       & $TM$       &          & $Omod$   \\ \hline \hline
$E_{1}$    & 2.0        & 2.58     & $\omega$    & 0.19   \\ 
$W_0$      & 2.3       & 3.56     & $W_0$         & 0.65   \\ 
$W_1$      & 1.5       & 1.19     & $W_1$         & 0.73   \\ 
$\lambda$ & 0.065   & 0.19     & $E_1$          & 1.68   \\ 
$\omega$  &     --      & 0.19     & $W_2$        & 0.58   \\ 
$\kappa$   &   --        & 0.19      & $W_3$        &  -0.5   \\ 
 &             &              & $\kappa$     & 0.19   \\ 
                  &             &              &  $\lambda$  & 0.61   \\
$m^{-1}$  & 1.43$\times10^{-3}$ & 28.06$\times10^{-4}$ & $m^{-1}$ & 2.8$\times10^{-2}$
\end{tabular}
\label{table:models}
\end{table}

\subsection{System-Bath Couplings and Spectral Lineshape}
In order to compare to experimentally measured spectra we take into account the interaction between the systems and their environment following the approach of Stock et al. \cite{Stock2005} in writing the coupling between the system and a collection of harmonic modes accounting for the effects of the protein pocket and the vibrational degrees of freedom of retinal not included in the system Hamiltonian:
\begin{eqnarray}
H_\mathrm{s,b}=\sum_\mathrm{b} S_\mathrm{b}\cdot\sum_k g_{\mathrm{b},k}(b^\dagger_{\mathrm{b},k}+b_{\mathrm{b},k})
\end{eqnarray}
where $S_\mathrm{b}$ is a system operator,  $g_{\mathrm{b},k}$ is the coupling strength of the $k$th mode of bath b, and $b^\dagger$ ($b$) is the corresponding creation (annihilation) operator of the bath mode.

The forms of $S_\mathrm{b}$ and the spectral density $J_\mathrm{b}(\omega')=\frac{\pi}{2}\sum_kg_{\mathrm{b},k}^2\delta(\omega'-\omega'_{\mathrm{b},k})$ depend on the model. We adopt the convention of Stock et al. \cite{Stock2005} and take $S_\phi=(1-\cos\phi)|1\rangle\langle1|$ for all three system models and $S_x=x|1\rangle\langle1|$ for the \textit{TM} and \textit{Omod} models, and Ohmic bath spectral densities $J_\mathrm{b}(\omega')=\eta_\mathrm{b}\omega'\exp(-\omega'/\omega'_{\mathrm{b}})$ with parameters given in Table~\ref{table:sb}.

\begin{table}[h]
 \caption{The parameters for the system-bath coupling terms adopted in the calculation.}
\begin{tabular}{c | c | c | c }
          & $SM$      &  $TM$       &    $Omod$   \\ \hline \hline
$\eta_x$   & --     & 0.1     & 0.1      \\ 
$\eta_\phi$   & 0.15     & 0.15     & 0.15      \\ 
$\omega_x$   & --     & $\omega$     & $\omega$      \\ 
$\omega_\phi$   & $\sqrt{\frac{W_0}{2m}}$     & $\sqrt{\frac{W_0}{2m}}$     & $\sqrt{\frac{W_0+4W_1}{2m}}$      \\ 
\end{tabular}
\label{table:sb}
\end{table}

The effect of system-bath coupling is treated within the Markovian-Redfield framework, described elsewhere \cite{Tscherbul2014JPCA}. Since we are primarily interested in spectroscopic observables at the stationary state, we further employ the Bloch-secular approximation, decoupling the dynamics of populations and coherences in the system energy eigenbasis. The former can be expressed as the solution to the master equation $\dot{P}=W\times P$, where $P_i(t)$ is the population of system energy eigenstate $i$ and $W_{ij}$ is the scattering rate from states $j$ to $i$ due to the system-bath coupling. Under the secular approximation, the absorption spectrum is simply the superposition of all bright states broadened by a Lorentzian lineshape:
\begin{eqnarray}
A_x(\epsilon)=\sum_i f_{x,i}\cdot\frac{|W_{ii}|}{(\epsilon-\epsilon_i)^2+W_{ii}^2}
\label{eqn:Absorption}
\end{eqnarray}
where $f_{x,i}=|\langle i|\hat{\mu}|x\rangle|^2$ is the oscillator's strength from state $x$ (the ground state in the \textit{cis}- or \textit{trans}-wells) to state $i$ with energy $\epsilon_i=\langle i|H_\mathrm{s}|i\rangle$,  and $W_{ii}=-\sum_jW_{ij}$ is the diagonal term of the rate matrix ensuring the conservation of population under the Bloch-secular approximation. We note that while the decoupling of population dynamics with that of the coherences employed by the Bloch-secular approximation generally introduces error in the transient regime, we are primarily interested in the stationary state where it yields reliable results.\\

\subsection{Spontaneous Emission Spectra}
We also examine the fluorescence emission spectra arising from the radiative decay of the excited state. To this end, in the master equation $\dot{P}=W\times P$ we also consider the rate kernel corresponding to a photon bath with $J^\mathrm{em}(\omega')=4\pi^2\omega'^3/3\epsilon_0(2\pi c)^3$, the photon spectral density from Planck's law where $\epsilon_0$ is the vacuum permittivity and $c$ is the speed of light, coupled to the system \cite{dodin2016}. The spontaneous emission spectrum of a given reduced system  density matrix at frequency $\omega'$ is obtained by collecting temperature-independent terms in the corresponding rate at $\omega' \approx-\omega'_{kl}$. In other words, the intensity of the emission spectrum at a certain frequency is proportional to the magnitude of the generalized rates connecting states with that frequency difference. For Bloch-secular approximated dynamics, the spectrum is given by
\begin{eqnarray}
E(\omega',t)\propto \left|\sum_{ij}W^\mathrm{em}_{ij}P_{j}(t)\delta(\omega'_{ij}+\omega')\right|
\end{eqnarray}
and the cumulative spectrum is,
\begin{eqnarray}
E(\omega')=\int_0^\infty dt~E(\omega',t).
\label{eqn:Emission}
\end{eqnarray}
In practice the $\delta$ function is replaced by a window function with a width $\Delta\epsilon$ on the order of the average nearest-neighbor energy gap of the system. Here we use the rectangular function $\delta(\omega'_{ij}+\omega')\rightarrow\Theta(\omega'_{ij}+\omega')\Theta(-\omega'_{ij}-\omega'-\Delta\omega')$. 

The three experimental emission spectra we compare with were obtained by exciting the rhodopsin sample with a continuous wave laser at various frequencies \cite{Mathies_Emission_exp}. Such an excitation condition is mimicked in our simulations by narrow band incoherent excitation coupling to the photon bath. 
Here only the stimulated absorption and emission terms of the corresponding Redfield rate kernel are included, and the spectral density is masked with a window function $J^{'\mathrm{em}}(\omega')=J^\mathrm{em}(\omega')D(\omega';\omega'_e,\delta\omega')$ centered at the excitation energy $\omega'_e$ and a finite width $\delta\omega'$. We take a Gaussian window function with $\delta\omega'=140$ cm$^{-1}$. The secular Redfield master equation with this source term of a given excitation energy is propagated from the ground state ($P(0)=|1\rangle$) to the steady state and Eq.~(\ref{eqn:Emission}) is used to extract the emission spectrum.

In addition, to account for the inhomogeneous broadening effect and better compare to the experiments, we further convolve Eqs.~(\ref{eqn:Absorption}) and (\ref{eqn:Emission}) with a Gaussian of width  2000 cm$^{-1}$.

\section{\label{sec:results} Results}
To gain insight into the relative merits of each of the three retinal models, we obtain the PF, optimizing first with respect to subsets of: the energy storage, three emission spectra excited at 472.7, 514.5, and 568.2 nm, and the maximum absorption frequency for the $cis$ and $trans$ conformers. The experimental values are reported in Table \ref{table:results_cleanmodels}.  
For the emission spectra we used the experimental data from Ref. \cite{Mathies_Emission_exp}.

\begin{figure}[h!]
\centering
\includegraphics[width=0.3\textwidth]{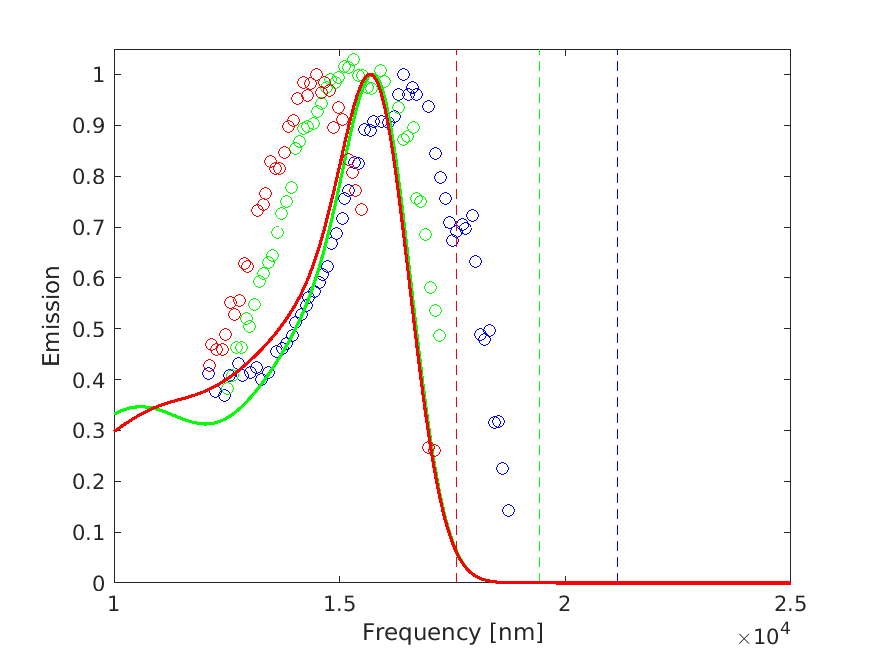}
\includegraphics[width=0.3\textwidth]{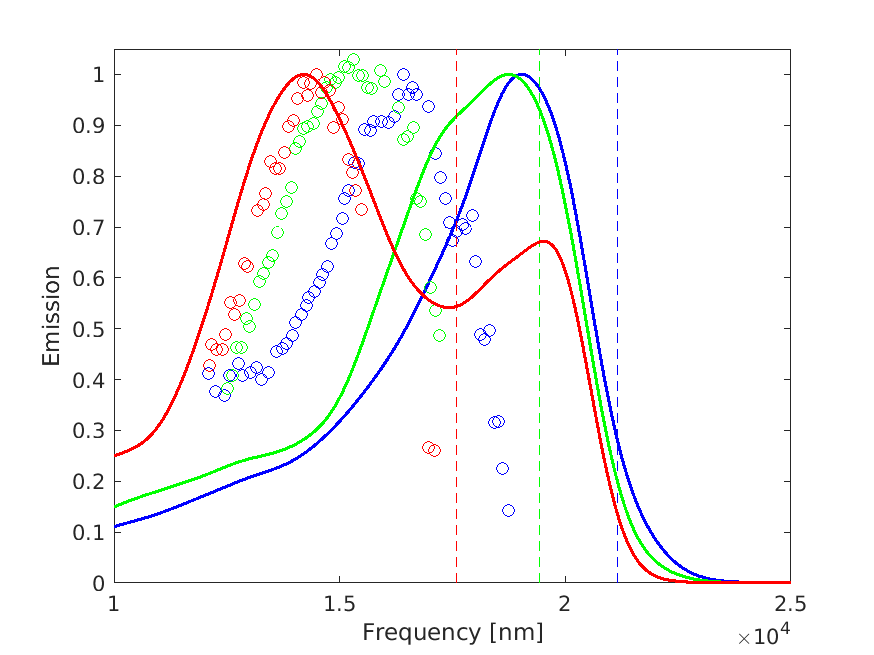}
\includegraphics[width=0.3\textwidth]{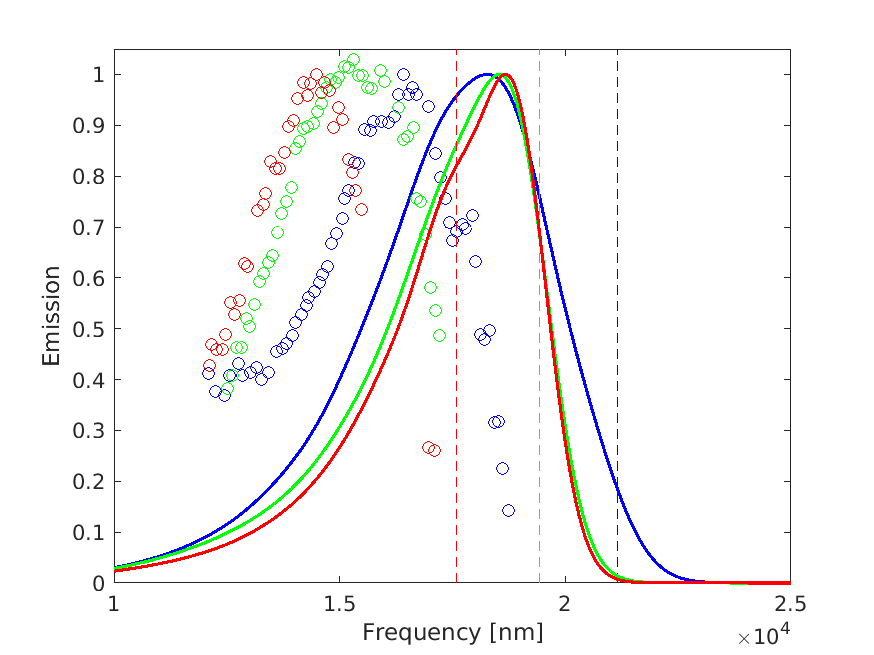}
\caption{Emission spectra predicted with the $SM$ model (upper panel), $TM$ model (middle panel), and the $Omod$ model (lower panel) using the parameters reported in Table \ref{table:models}. The symbols represent the experimental emission spectra excited at the three frequencies \cite{Mathies_Emission_exp}, indicated by the vertical dashed lines, and the solid curves are the predicted emission spectra.}
\label{fig:clean models_EmiSpectra}
\end{figure}

To quantify the accuracy of each model with respect to the emission spectrum we consider the sum square error for each emission spectrum, $f(E_{i})$,
\begin{eqnarray}
f(E_{i}) = \sum_i  \Big( g(\nu) - \hat{g}(\nu)  \Big)^2,
\end{eqnarray}
where  $g(\nu)$ is the calculated emission spectra at frequency $\nu$, and $\hat{g}(\cdot)$ are the corresponding experimental values of the emission spectra from Ref. \cite{Mathies_Emission_exp}. $E_i$ are the excitation frequencies $[472.7, 514.5, 568.2]$ in nm from Ref.  \cite{Mathies_Emission_exp}, that yield the three observed emission spectra.
We quantified the accuracy of the models for the energy storage, and the maximum absorption frequencies for the $cis$ and $trans$ states using their  absolute error.
For all three models, the MOBOpt algorithm learned the PF as a function of  $\Delta\boldsymbol\theta_{i}$, a correction factor for each parameter from Table \ref{table:models}; $\boldsymbol\theta_{i}  = \boldsymbol\theta^{0}_{i} + \Delta\boldsymbol\theta_{i}$. 
 For the $TM$ model we did not include the $\omega$ parameter in the search procedure, and for all models the mass parameter in the kinetic energy operator was held fixed.
In Table  \ref{table:results_cleanmodels} we report the calculated values for all three models when $\Delta\boldsymbol\theta_{i} = 0$ (i.e., the models in the literature).
The corresponding emission spectra are displayed in Figure \ref{fig:clean models_EmiSpectra}.
All are clearly in poor agreement with the experimental data, one of the motivations for an effort to improve the models.\\

For all calculations we set the number of Pareto points to 100, and in the starting of the optimization we randomly sample 50 different $\Delta\boldsymbol{\theta}$.
By iteratively sampling new points with Eq. \ref{eqn:mobopt_aqn} we can update each objective function and reconstruct the PF using the NSGAII algorithm \cite{NSGA-II}.
The maximum number of optimizations for each example was set to 200 to ensure proper exploration of the parameter space. Additionally, one sanity check was to verify that the proposed $Y_{P}$ are positive, since for this work all error functions are positive.\\

\begin{table}[h!]
 \caption{The calculated values of the energy storage, and $cis$ and $trans$ states maximum absorption frequencies for all three different models with parameters from Table \ref{table:models}, as well as the experimental values for energy storage \cite{EnergyStorage_rhodopsin}, and the maximum absorption frequencies for $cis$ and $trans$ \cite{Abs_rhodopsin}. }
\begin{tabular}{c | c c c  | c}
               & $SM$    &  $TM$   & $Omod$   & Exp.   \\ \hline \hline
Energy Storage [eV] & 0.496 & 1.26   & 1.3    & 1.39   \\
max $cis$  [nm]    & 623.61 & 490.37 & 497.34 & 504.85 \\
max $trans$ [nm]   & 654.21 & 530.30 & 611.35 & 538.86
\end{tabular}
\label{table:results_cleanmodels}
\end{table}

\subsection{\label{sec:PF_Emi_ES} Pareto Front for Emission and Energy Storage}
\begin{figure}[h!]
\centering
\includegraphics[width=0.3\textwidth]{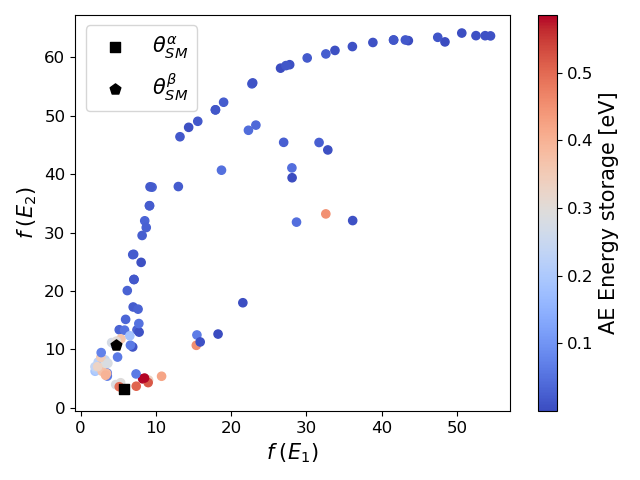}
\includegraphics[width=0.3\textwidth]{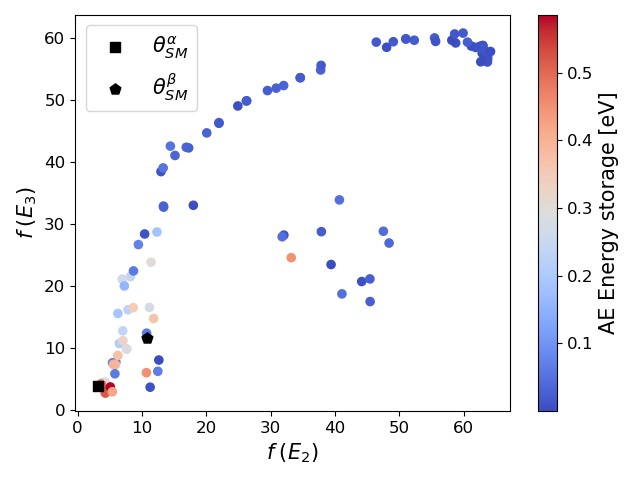}
\includegraphics[width=0.3\textwidth]{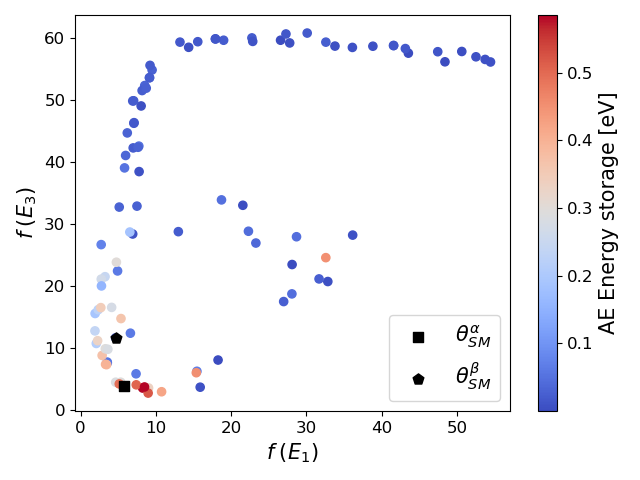}
\caption{PF for the $SM$ model computed with respect to the three emission spectra and the energy storage. The energy storage accuracy was quantified with the absolute error. The black symbols indicate the selected two models $\boldsymbol{\theta}^{\alpha}_{SM}$ and $\boldsymbol{\theta}^{\beta}_{SM}$.}
\label{fig:PF_SM_Emi_Abs}
\end{figure}

\begin{figure}[h!]
\centering
\includegraphics[width=0.3\textwidth]{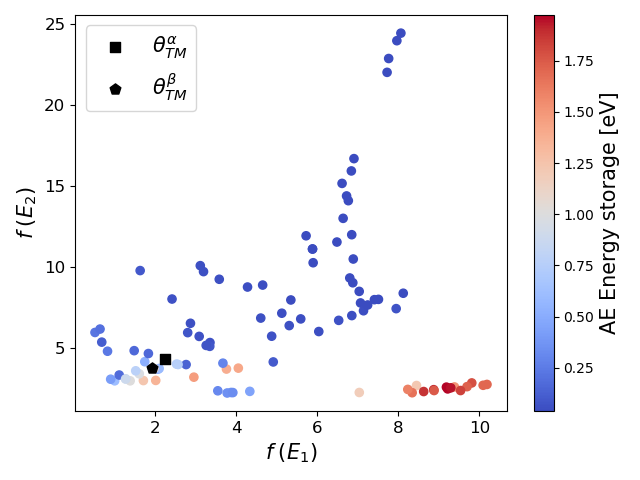}
\includegraphics[width=0.3\textwidth]{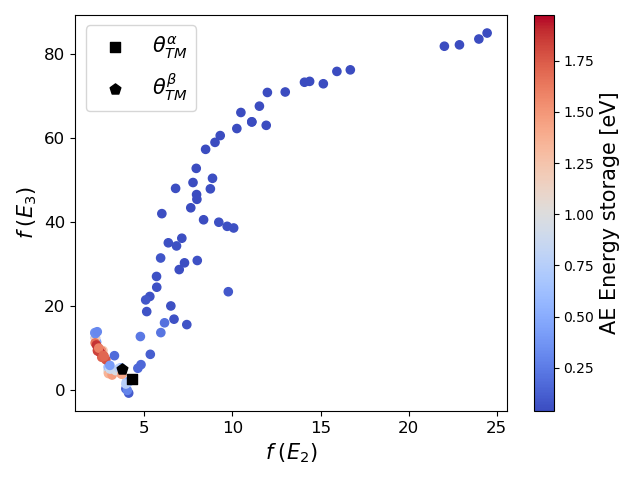}
\includegraphics[width=0.3\textwidth]{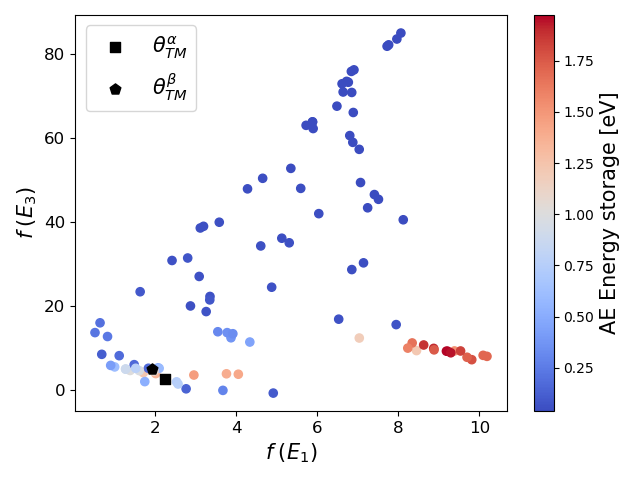}
\caption{PF for the $TM$ model computed with respect to the three different emission spectra and the energy storage. The black symbols indicate the selected two models $\boldsymbol{\theta}^{\alpha}_{TM}$ and $\boldsymbol{\theta}^{\beta}_{TM}$. Note that the axes of the three panels are not on the same scale.
}
\label{fig:PF_TM_Emi_Abs}
\end{figure}

\begin{figure}[h!]
\centering
\includegraphics[width=0.3\textwidth]{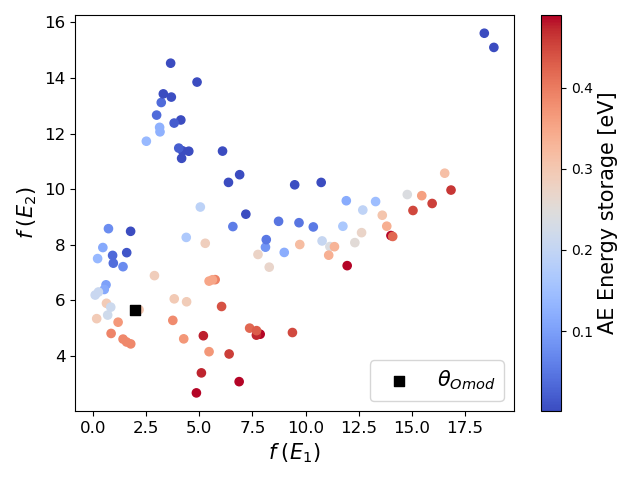}
\includegraphics[width=0.3\textwidth]{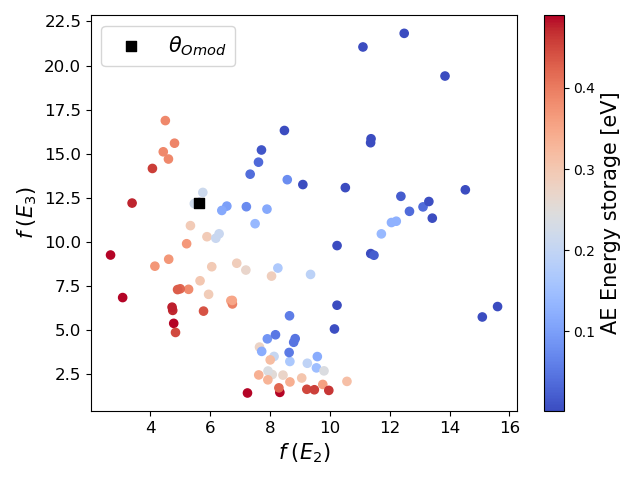}
\includegraphics[width=0.3\textwidth]{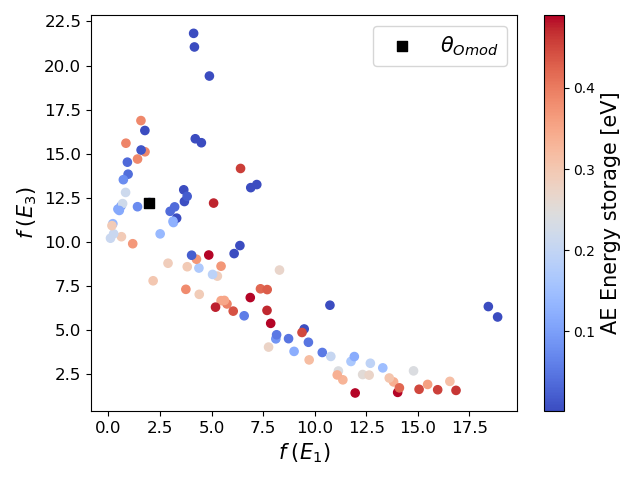}
\caption{PF for the $Omod$ model computed with respect to the three different emission spectra and the energy storage. The black symbol indicates the selected model $\boldsymbol{\theta}_{Omod}$. Note that the axes of the three panels are not on the same scale.
}
\label{fig:PF_Oliv_Emi_Abs}
\end{figure}

In this section we present the PF, considering only the experimental subset of the three emission spectra and the energy storage. 
Figure \ref{fig:PF_SM_Emi_Abs} displays the $Y_{P}$ computed with the $SM$ model, Figure \ref{fig:PF_TM_Emi_Abs} for the $TM$ model, and Figure \ref{fig:PF_Oliv_Emi_Abs} for the $Omod$ model. All figures consist of three panels that correspond to two dimensional projections of the full four dimensional space of $\mathbf{F}$.

\begin{figure}[h!]
\centering
\includegraphics[width=0.3\textwidth]{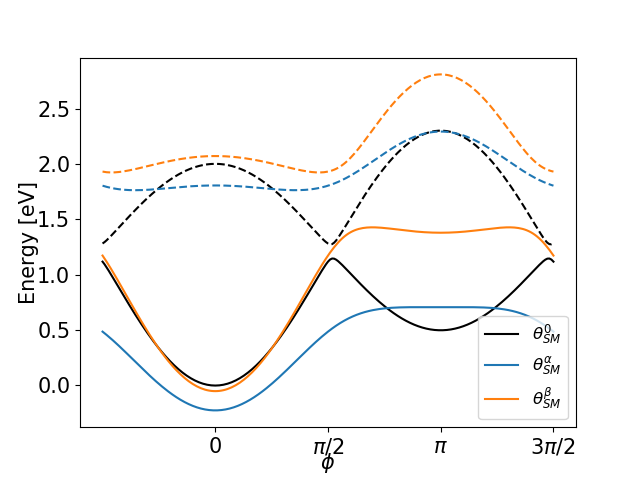}
\includegraphics[width=0.3\textwidth]{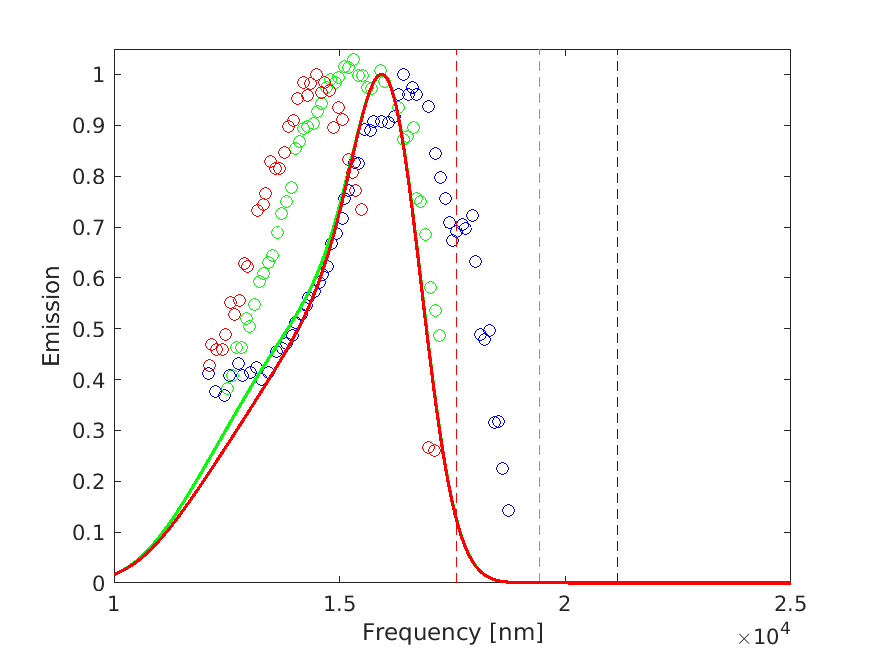}
\includegraphics[width=0.3\textwidth]{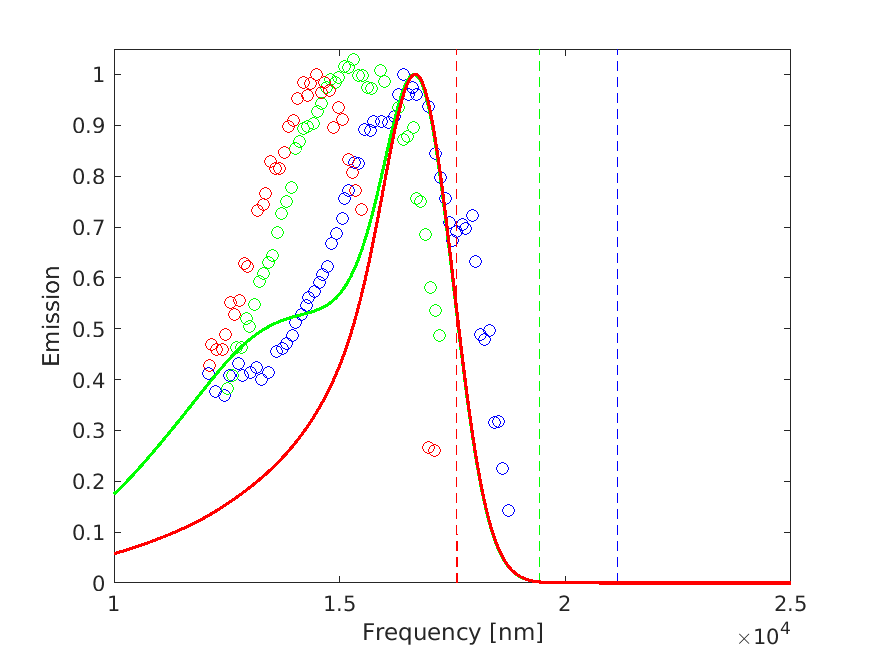}
\caption{(upper panel) Adiabatic surfaces for the original $SM$ model $\boldsymbol{\theta}_{SM}^{0}$ and those of $\boldsymbol{\theta}_{SM}^{\alpha}$ and $\boldsymbol{\theta}_{SM}^{\beta}$. (middle pannel) Predicted emission spectra of $\boldsymbol{\theta}_{SM}^{\alpha}$ model. (lower panel) Predicted emission spectra of $\boldsymbol{\theta}_{SM}^{\beta}$ model. We adopt the same convention as in Figure \ref{fig:clean models_EmiSpectra}.}
\label{fig:PF_SM_EmiSpectra_Emi_ES}
\end{figure}

The PF for the $SM$ model shows that the emission spectra and the energy storage are not simultaneously minimized. Two values of $\boldsymbol\theta$ are identified, $\boldsymbol\theta^\alpha_{SM} = [1.58, 1.97, 0.55, 0.64]$ and  $\boldsymbol\theta^\beta_{SM} = [2.02, 2.73, 0.56, 0.33]$ corresponding to the lowest mean error for the three emission spectra, and the lowest energy storage error and mean emission spectra, respectively. 
 $\boldsymbol\theta^\alpha_{SM}$ shows an energy storage of 0.92 eV, and 1.42 eV for $\boldsymbol\theta^\beta_{SM}$, compared to the experimental value of 1.39 eV. Both are improved compared to the value in Table \ref{table:results_cleanmodels}. Figure \ref{fig:PF_SM_EmiSpectra_Emi_ES} shows the emission spectra and the adiabatic potential energy surfaces for both $SM$ models.
 Both set of spectra are in poor agreement with experiment.
 
\begin{figure}[h!]
\centering
\includegraphics[width=0.3\textwidth]{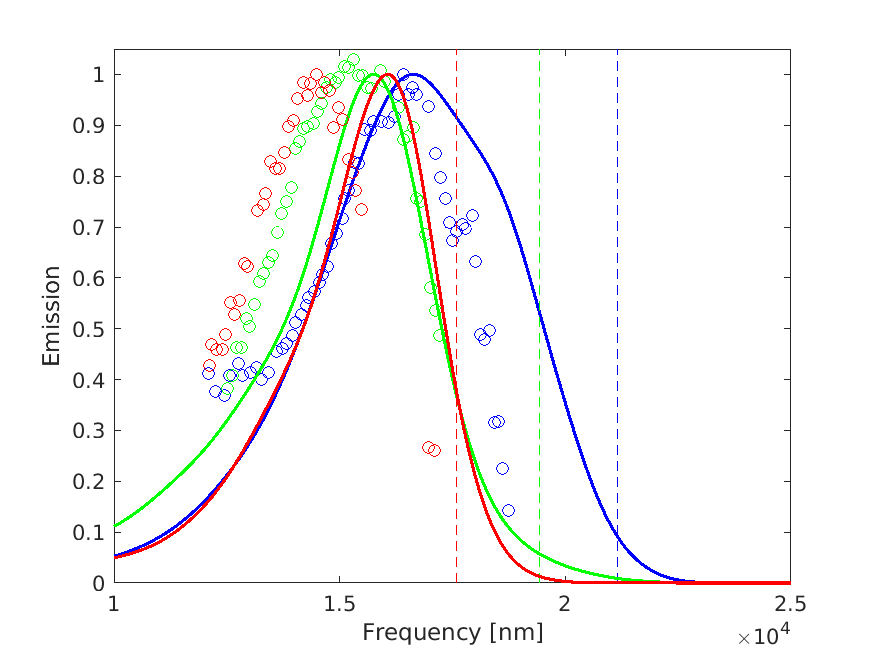}
\includegraphics[width=0.3\textwidth]{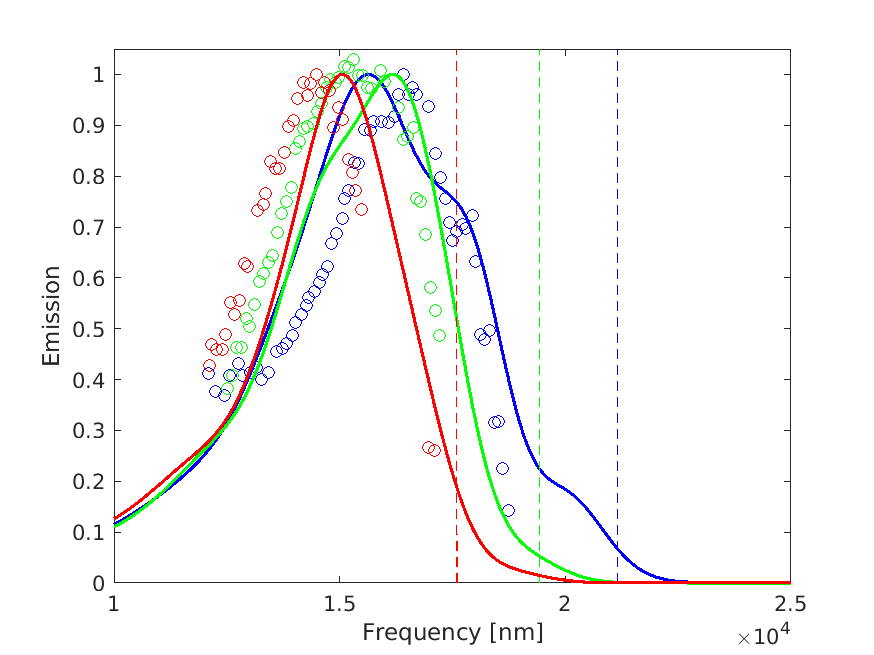}
\caption{Emission spectra predicted with $TM$  model with various parameter sets. (upper panel) $\boldsymbol{\theta}^{\alpha}_{TM}$, and (lower panel) $\boldsymbol{\theta}^{\beta}_{TM}$. We adopt the same convention as in Figure \ref{fig:clean models_EmiSpectra}.}
\label{fig:PF_TM_EmiSpectra_Emi_ES}
\end{figure}

\begin{figure}[h!]
\centering
\includegraphics[width=0.32\textwidth]{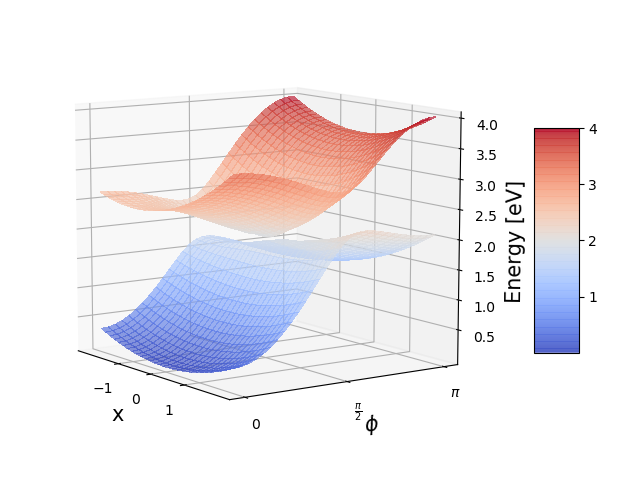}
\includegraphics[width=0.32\textwidth]{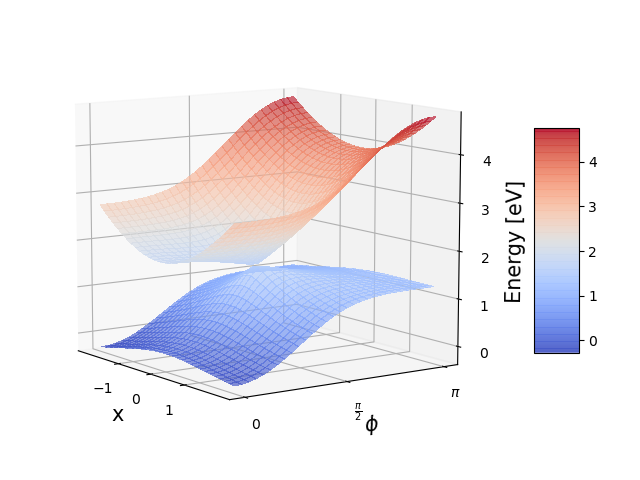}
\includegraphics[width=0.32\textwidth]{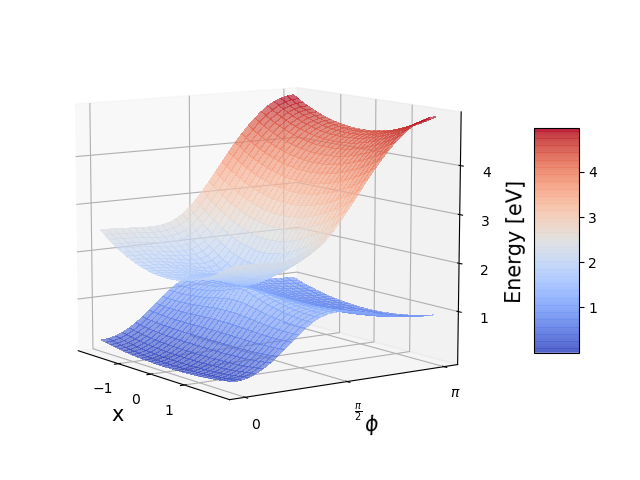}
\caption{Adiabatic surfaces for the selected $TM$ models. (upper panel) Parameters from Table \ref{table:models}, (middle panel) $\boldsymbol{\theta}^{\alpha}_{TM}$, and (lower panel) $\boldsymbol{\theta}^{\beta}_{TM}$.
}
\label{fig:TM_PES}
\end{figure}

Figure~\ref{fig:PF_TM_Emi_Abs} displays the PF for the $TM$ model. The visual impression of points gathering towards the lower left corner implies that reasonably good agreement with experiment is possible with this model. The selected points in black, corresponding to $\boldsymbol\theta^\alpha_{TM} = [1.88, 3.87, 0.641, 0.074, 0.627]$ and  $\boldsymbol\theta^\beta_{TM} = [1.833,  4.477,  1.143, -0.0158,  0.351]$ give energy storage of 1.28 and 0.69 eV respectively. Whereas the emission spectra for $\boldsymbol\theta^\beta_{TM}$ is in good agreement with experiment, as shown in the lower panel of Figure \ref{fig:PF_TM_EmiSpectra_Emi_ES}, its energy storage is half of the experimental value. By contrast, $\boldsymbol\alpha^\alpha_{TM}$ gives very good agreement with the experimental energy storage, but is not in good agreement with the emission spectra (upper panel, Figure \ref{fig:PF_TM_EmiSpectra_Emi_ES}). 
Adiabatic potential energy surfaces for the three $TM$ models are shown in Figure \ref{fig:TM_PES}. Clearly there are significant differences in the structure of these surfaces and the relative positioning of the conical intersection. 

\begin{figure}[h!]
\centering
\includegraphics[width=0.4\textwidth]{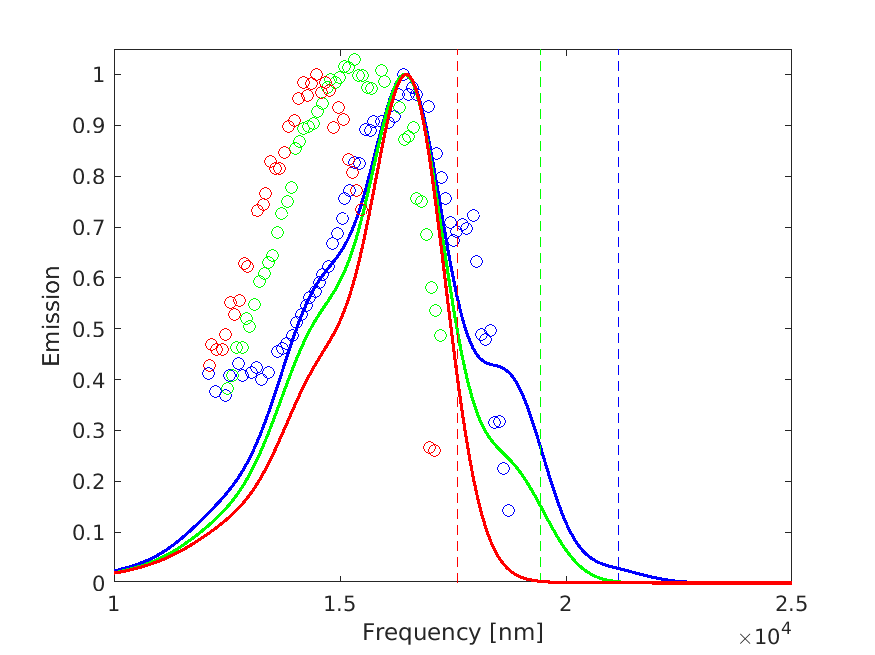}
\caption{Emission spectra predicted with the $Omod$ model using the $\boldsymbol{\theta}_{Omod}$ parameters given in the text. We adopt the same convention as in Figure \ref{fig:clean models_EmiSpectra}.}
\label{fig:Oliv_EmiSpectra_Emi_ES}
\end{figure}

Figure  \ref{fig:PF_Oliv_Emi_Abs} shows the PF for the $Omod$ model. The lower panel is particularly interesting because the PF avoided the lower left corner, indicating that simultaneous agreement of these properties to experiments is not possible. For the $Omod$ model, the parameter set that best describes all emission spectra with an accurate energy storage (1.36 eV),  is given by $\boldsymbol{\theta}_{Omod} = [ 0.2709,  0.6802,  0.5314, 1.5213,  0.2066,  0.5498, -0.636,  0.6623]$, as shown in Figure  \ref{fig:Oliv_EmiSpectra_Emi_ES}.
While the energy storage is in excellent agreement, the emission spectra are poor.

\begin{figure}[h!]
\centering
\includegraphics[width=0.3\textwidth]{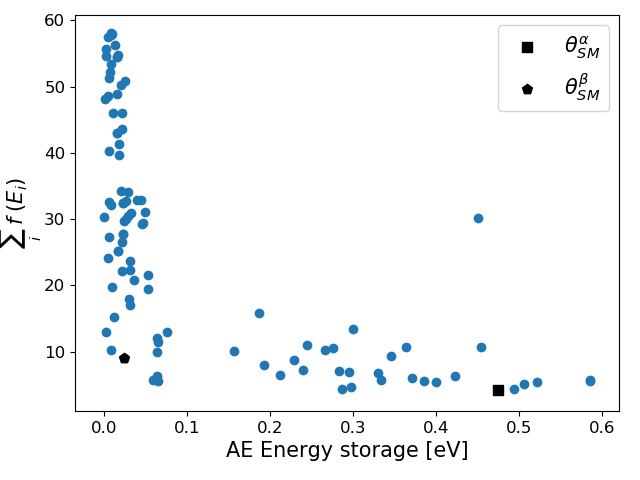}
\includegraphics[width=0.3\textwidth]{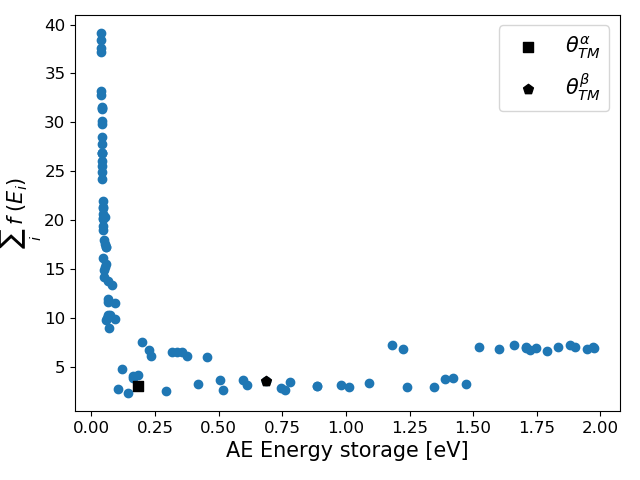}
\includegraphics[width=0.3\textwidth]{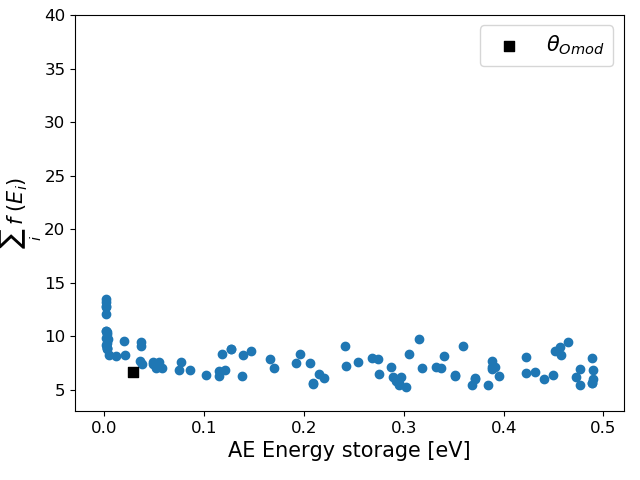}
\caption{PF for the average square error for the different emission spectra and the absolute error of the energy storage computed with the (upper panel)  $SM$ model, (middle panel) $TM$ model, and (lower panel) $Omod$ model. Note that the axes of the three panels are not on the same scale.}
\label{fig:PF_allmodels_EmiSpectra_Emi_Abs}
\end{figure}

The difficulty of interpreting the PF data motivated us to simplify matters by computing the mean square error for the emission spectra and the energy storage absolute error with the PF for each model. 
From Figure \ref{fig:PF_allmodels_EmiSpectra_Emi_Abs} we can observe that the $TM$ and $Omod$ models have a smaller error in the predicted emission spectra than the $SM$ model. This approach motivates the study in the next section.

\subsection{\label{sec:PF_meanEmi_Abs_ES} Pareto Front for mean Emission, Absorption and Energy Storage}
In the previous section we showed that the mean error for the emission spectra is a useful metric to learning a set of viable parameters.  Here, we use $\sum_{i=1}^{3}f(E_i)$ as one of the elements of $\mathbf{F}$ to quantify the accuracy of all three emission spectra. Moreover, we expand the number of target observables to include the peak absorption frequencies for the $cis$ and $trans$ conformers so that $\mathbf{F} = \Big[\sum_{i=1}^{3}f(E_i), |\nu_{cis} - \hat{\nu}_{cis}|,|\nu_{trans} - \hat{\nu}_{trans}|, |ES - 1.39|\Big]$, where $\hat{\nu}$'s are the frequencies reported in Table \ref{table:results_cleanmodels}, and $ES$ is the calculated energy storage.
We follow the same procedure and use the MOBOpt algorithm to look for the PF for these four objectives.

The PF for the $SM$ model was computed but was found to be incapable of jointly describing these observables, not even individually.  The lowest error value for the $cis$ and $trans$ absorption frequencies is relatively large. 
Given this result, and the poor results reported in the previous section, the $SM$ model is not regarded as a ``contender'' for a usable description of the system.

\begin{figure}[h!]
\centering
\includegraphics[width=0.3\textwidth]{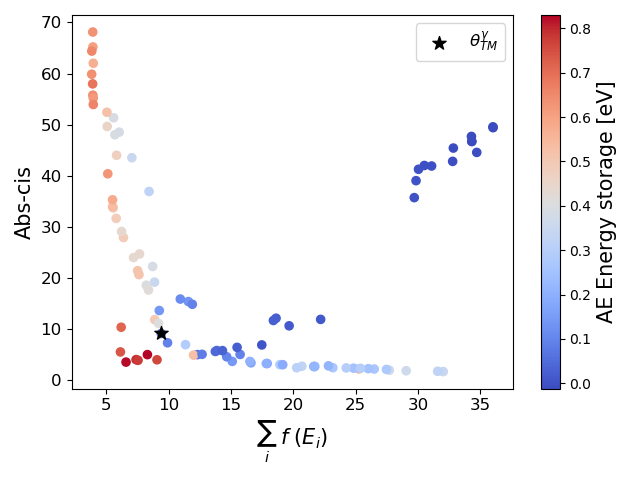}
\includegraphics[width=0.3\textwidth]{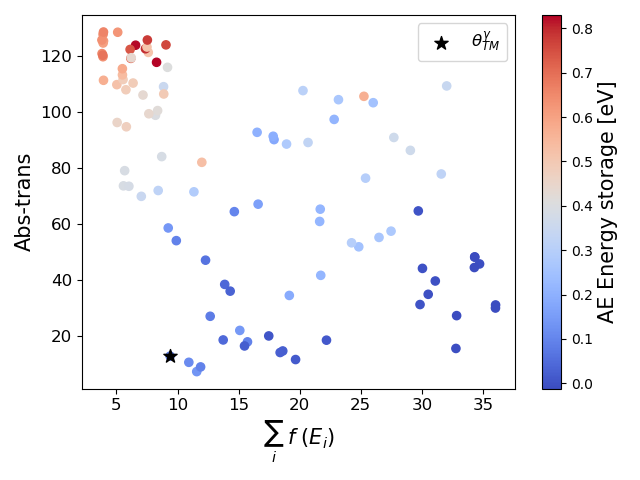}
\includegraphics[width=0.3\textwidth]{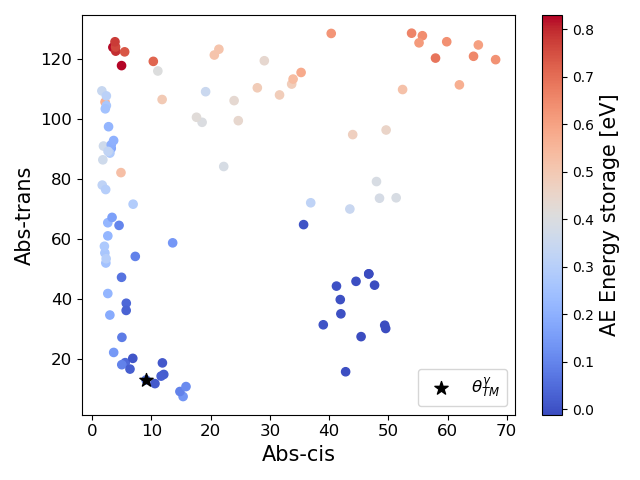}
\caption{PF for the $TM$ model computed with respect to the sum of three different emission spectra errors, absolute difference for the maximum absorption frequency  for the $cis$ and $trans$ states, and the absolute error for the energy storage.  Note that the axes of the three panels are not in the same scale.
}
\label{fig:PF_TM_meanEmi_Abs_ES}
\end{figure}

\begin{figure}[h!]
\centering
\includegraphics[width=0.4\textwidth]{fig3b}
\includegraphics[width=0.4\textwidth]{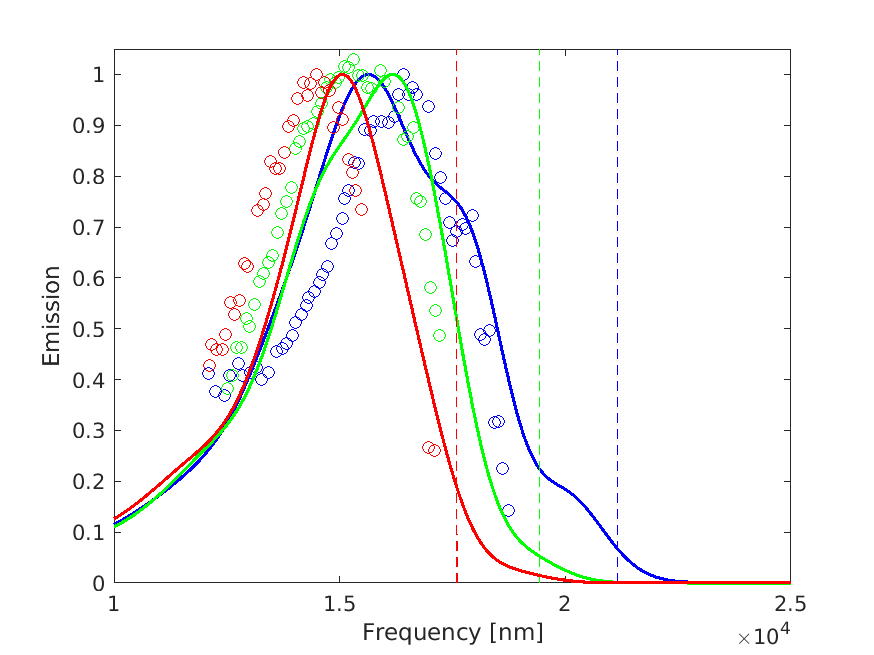}
\caption{Emission spectra predicted with $TM$ model with various parameter sets. (upper panel) Parameters from Table \ref{table:models}. (lower panel) $\boldsymbol{\theta}^{\gamma}_{TM}$. 
We adopt the same convention as in Figure \ref{fig:clean models_EmiSpectra}.}
\label{fig:TM_meanEmi_Abs_ES}
\end{figure}

The results for the $TM$ model are different. The PF, displayed in Figure \ref{fig:PF_TM_meanEmi_Abs_ES}, shows that this model is capable of better describing this set of physical observables.
The best values were obtained with $\boldsymbol{\theta}^{\gamma}_{TM} = [ 2.385,  3.382,   0.8418,  0.1987,  -0.00519]$. 
The predicted energy storage is 1.31 eV, and 516.6 and 515.7 nm are the max absorption frequencies for the $cis$ and $trans$ conformers, respectively. This compares to an experimental energy storage of 1.39 eV and max absorption at 504 nm and 539 nm, respectively.
Figure \ref{fig:TM_meanEmi_Abs_ES} lower panel displays the predicted emission spectra with these parameters, which are in far better agreement with experiment than the original $TM$ parameters.

\begin{figure}[h!]
\centering
\includegraphics[width=0.4\textwidth]{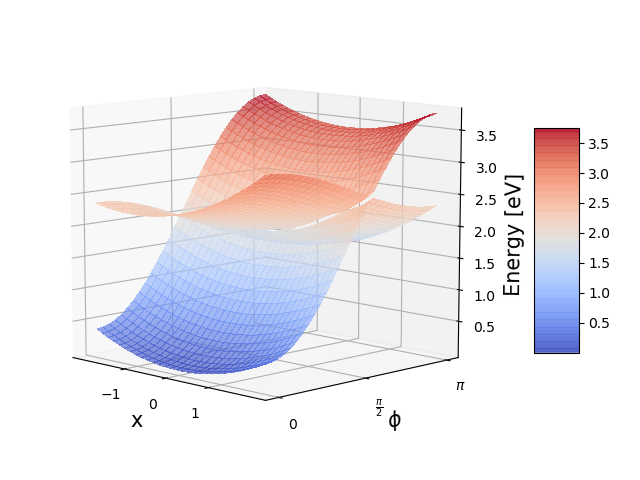}
\includegraphics[width=0.4\textwidth]{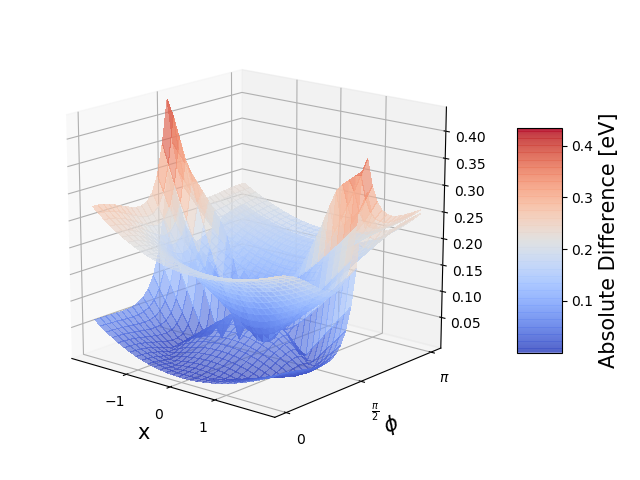}
\caption{(upper panel) Adiabatic surfaces for $\boldsymbol{\theta}^{\gamma}_{TM}$.
(lower panel) Absolute difference between the adiabatic surfaces produced with $\boldsymbol{\theta}^{\gamma}_{TM}$ and parameters from Table \ref{table:models}.}
\label{fig:TM_PES_meanEmi_Abs_ES}
\end{figure}

The biggest and perhaps most consequential difference introduced in the $\boldsymbol{\theta}^{\gamma}_{TM}$ model is its $\lambda$ parameter, the nonadiabatic coupling, that is nearly vanishing. This leads to the similarity between the adiabatic surfaces with their diabatic counterparts, as seen in Figure \ref{fig:TM_PES_meanEmi_Abs_ES}. In contrast to a ``peaked'' conical intersection typically seen in a $TM$ model, the energy gap between the two adiabatic surfaces opens up only slightly along the $|x|>0$ direction owing to the small $\lambda$ value. Consequently, the gradient of the adiabats is dominantly aligned in the $\phi$ direction, especially in regions close to the Franck-Condon states expected to be most emissive. In such a case one expects a relaxation trajectory that retains more memory of its initial starting point, leading to a stronger dependence of emission spectrum on the excitation frequency as observed in the experiment. We note that the situation in the actual rhodopsin is much more complex and might involve more degrees of freedom afforded by an effective two-state two-model model.\cite{Olivucci_model} However, our analysis of the $\boldsymbol{\theta}^{\gamma}_{TM}$ model which produces good agreement with experimental emission spectra sheds light on the relevant energy landscape. 

\section{\label{sec:summary} Summary}
There are two aspects to the results in this paper, the nature of the multi-objective optimization procedure, and the application to the retinal models.
In the first category we note that the optimization of physical models to reproduce a  set of target observables is a common task where machine-learning algorithms have proven to be useful.
When multiple target observables are considered, the standard procedure is to linearly combine all error functions. This can lead to models with a variety of accuracy for the target observables due to possible existence of multiple local minima. 
This type of optimization scheme will only succeed if there is a single set of parameters jointly optimizing all the target functions, which is a strong assumption. To overcome these obstacles we used a multi-objective optimization protocol where the goal is to learn the limits of each objective through the Pareto front.
The results illustrate that we can learn more about the robustness of a physical model by finding the Pareto front  than would results from using a single objective optimization scheme.
Furthermore, the MOBOpt algorithm used in this work relies on individual Gaussian processes for each target observable reducing number of evaluations that are required.

One of the key features of multi-objective optimization is the possibility of recognizing a joint minimizer shared across different objectives as displayed in the Pareto front in Figures \ref{fig:PF_SM_Emi_Abs},  \ref{fig:PF_TM_Emi_Abs}, and \ref{fig:PF_Oliv_Emi_Abs}. 
This is in contrast to the Pareto front for the $cis$ and $trans$ conformers' maximum absorption frequencies which displays a \emph{wall} shape, e.g., Figure \ref{fig:PF_TM_meanEmi_Abs_ES}, which indicates that the two functions do not share a minimizer in the parameter space.
Furthermore, given the flexibility of MOBOpt algorithm  and the efficacy of Gaussian processes, the present procedure is independent of the simulation procedure for each observable, and could be used for other problems besides quantum dynamics.

To utilize this algorithm, we considered the emission spectra excited at three different wavelengths, combined with the maximum absorption frequencies for the $cis$ and $trans$ conformers and the energy storage for the retinal chromophore in rhodopsin, all evaluated in the steady state.
This set of observables were simulated with three different models: $single-mode$, $two-mode$ models, and a third one based on Ref. \cite{Olivucci_model}, accounting for the two lowest energy electronic states.
We exemplify the character of the Pareto front by considering different combinations of all six observables.
For example, we demonstrate that the $SM$ model is incapable of fitting the emission spectra and the maximum absorption frequencies without compromising other observables.
However, for the $TM$ and the $Omod$ model the Pareto fronts show that more accurate model exists without jeopardizing the accuracy of the rest of the observables.

We found that the $TM$ model with the $\boldsymbol{\theta}^{\gamma}_{TM}$ parameters  $E_1=2.385$, $W_0=3.382$, $W_1=3.382$, $\omega = 0.8418$, $\kappa = 0.1987$, and $\lambda= -0.00519$  is the model that best describe all the different experimental observables considered in this study. 
Here the $\lambda$ value is significantly different from the literature value (Table \ref{table:models}); $E_1=2.58$, $W_0=3.56$, $W_1=1.19$, $\omega = 0.19$, $\kappa = 0.19$, and $\lambda= 0.19$,
Most significantly, the small magnitude of $\lambda$, the nonadiabatic coupling, of the $\boldsymbol{\theta}^{\gamma}_{TM}$ model is likely responsible for the good agreement with the experimental emission spectra. This is a trait that might also be applicable to more complicated, higher dimensional models to be investigated in the future.
Indeed the lack of quantitative agreement with experimental data in all the models motivates such studies, which are ongoing in our laboratory.

\begin{acknowledgments}
This work was supported by U.S. Air Force Office of Scientific Research (AFOSR) grant FA9550-20-1-0354.
\end{acknowledgments}

\section*{Data Availability Statement}
The data that support the findings of this study are available from the corresponding author upon reasonable request.




\end{document}